\RequirePackage{fix-cm}
\documentclass[smallextended]{svjour3}       % onecolumn (second format)
\smartqed  % flush right qed marks, e.g. at end of proof
\usepackage{graphicx}
\usepackage[english]{babel}
\usepackage[comma,authoryear]{natbib}
\usepackage[T1]{fontenc}       
\usepackage{eurosym}
\usepackage[latin1]{inputenc}  
\usepackage{footnote}
\usepackage[countmax]{subfloat}
\usepackage{mathtools}  
\usepackage{placeins}

\usepackage{amsmath}
\usepackage{tablefootnote}

 \journalname{Theory and Decision}

\begin{document}

\title{Confidence Biases and Learning among Intuitive Bayesians%\thanks{Grants or other notes
}


\author{Louis L\'evy-Garboua         \and
        Muniza Askari \and
        Marco Gazel
}


\institute{Louis L\'evy-Garboua \at
              Paris School of Economics, Universit\'e Paris 1 Pantheon-Sorbonne, and Centre d'Economie de la Sorbonne, 106-112 Bd de l'H\^opital 75013, Paris, France \\
              Tel.: +33 6 85 11 18 68\\
             \email{louis.levy-garboua@univ-paris1.fr}           %  \\
           \and
           Muniza Askari \at
              Centre d'Economie de la Sorbonne\\
               \email{muneeza\_askari@hotmail.com} 
              \and
             Marco Gazel\at
             Paris School of Economics, Universit\'e Paris 1 Panth\'eon Sorbonne and Centre d'Economie de la Sorbonne, 106-112 Bd de l'H\^opital - 75013 Paris\\
             \email{ marco-antonio.gazel@univ-paris1.fr}      
}

\date{Received: date / Accepted: date}

\maketitle

\begin{abstract}
We design a double-or-quits game to compare the speed of learning one's specific ability with the speed of rising confidence as the task gets increasingly difficult. We find that people on average learn to be overconfident faster than they learn their true ability and we present an Intuitive-Bayesian model of confidence which integrates confidence biases and learning. Uncertainty about one's true ability to perform a task in isolation can be responsible for large and stable confidence biases, namely limited discrimination, the hard-easy effect, the Dunning-Kruger effect, conservative learning from experience and the overprecision phenomenon (without underprecision) if subjects act as Bayesian learners who rely only on sequentially perceived performance cues and contrarian illusory signals induced by doubt. Moreover, these biases are likely to persist since the Bayesian aggregation of past information consolidates the accumulation of errors and the perception of contrarian illusory signals generates conservatism and under-reaction to events. Taken together, these two features may explain why intuitive Bayesians make systematically wrong predictions of their own performance.

\keywords{Confidence biases \and intuitive-Bayesian \and learning \and double or quits experimental game \and doubt \and contrarian illusory signals}
\end{abstract}

\section{Introduction}

In many circumstances, people appear to be "overconfident" in their own abilities and good fortune. This may occur when they compare themselves with others, massively finding themselves "better-than-average" in familiar domains (eg., \citealt{svenson1981}, \citealt{kruger1999}), when they overestimate their own absolute ability to perform a task (eg., \citealt{lichtenstein1977}, \citealt{lichtenstein1982}), or when they overestimate the precision of their estimates and forecasts (eg., \citealt{oskamp1965}). \cite{moore2008} designate these three forms of overconfidence respectively as overplacement, overestimation, and overprecision. We shall here be concerned with how people overestimate, or sometimes underestimate, their own absolute ability to perform a task in isolation. Remarkably, however, our explanation of the estimation bias predicts the overprecision phenomenon as well.

The estimation bias refers to the discrepancy between ex post objective performance (measured by frequency of success in a task) with ex ante subjectively held confidence \citep{lichtenstein1982}. It has first been interpreted as a cognitive bias caused by the difficulty of the task (e.g.,\citealt{griffin1992}). It is the so called "hard-easy effect" \citep{lichtenstein1977}: people underestimate their ability to perform an easy task and overestimate their ability to perform a difficult task. However, a recent literature has challenged this interpretation by seeking to explain the apparent over/underconfidence by the rational-Bayesian calculus of individuals discovering their own ability through experience and learning \citep{moore2008, grieco2009, benoit2011, van2011}. While the cognitive bias view describes self-confidence as a stable trait, the Bayesian learning perspective points at the experiences leading to over- or under-confidence. The primary goal of this paper is to propose a parsimonious integration of the cognitive bias and the learning approach.

We design a real-effort experiment which enables us to test the respective strengths of estimation biases and learning. People enter a game in which the task becomes increasingly difficult -\textit{i.e.} risky- over time. By comparing, for three levels of difficulty, the subjective probability of success (confidence) with the objective frequency at three moments before and during the task, we examine the speed of learning one's ability for this task and the persistence of overconfidence with experience. We conjecture that subjects will be first underconfident when the task is easy and become overconfident when the task is getting difficult. However, "difficulty" is a relative notion and a task that a low-ability individual finds difficult may look easy to a high-ability person. Thus, we should observe that overconfidence declines with ability and rises with difficulty. The question raised here is the following: if people have initially an imperfect knowledge of their ability and miscalibrate their estimates, will their rising overconfidence as the task becomes increasingly difficult be offset by learning, and will they learn their true ability fast enough to stop the game before it is too late? 

The popular game \textit{"double or quits"} fits the previous description and will thus inspire the following experiment. A modern version of this game is the world-famous TV show \textit{"who wants to be a millionaire"}. In the games of \textit{"double or quits"} and \textit{"who wants to be a millionaire"}, players are first given a number of easy questions to answer so that most of them win a small prize. At this point, they have an option to quit with their prize or double by pursuing the game and answering a few more questions of increasing difficulty. The same sort of double or quits decision may be repeated several times in order to allow enormous gains in case of repeated success. However, if the player fails to answer one question, she must step out of the game with a consolation prize of lower value than the prize that she had previously declined.

Our experimental data reproduces the double or quits game. We observe that subjects are under-confident in front of a novel but easy task, whereas they feel overconfident and willing to engage in tasks of increasing difficulty to the point of failing.

We propose a new model of "intuitive Bayesian learning" to interpret the data and draw new testable implications. Our model builds on ideas put forward by \citet{erev1994} and \citet{moore2008}. It is Bayesian like \citet{moore2008}, while viewing confidence as a subjective probability of success, like \citet{erev1994}. However, it introduces intuitive rationality to overcome a limitation of  the rational-Bayesian framework which is to describe how rational people learn from experience without being able to predict the formation of confidence biases \textit{before} completion of a task. This is not an innocuous limitation because it means, among other things, that the rational-Bayesian theory is inconsistent with the systematic probability distortions observed in decisions under risk or uncertainty since the advent of prospect theory \citep{kahneman1979prospect}. Therefore, we need to go deeper into the cognitive process of decision.  Subjects in our view derive their beliefs exclusively from their prior and the informative signals that they receive. However, "intuitive Bayesians" decide on the basis of the sensory evidence that they perceive sequentially. If they feel uncertain of their prior belief, they will perceive the objection to it triggered by their doubt and wish to "test" its strength before making their decision, like those decision makers weighting the pros and cons of an option. The perceived objection to a rational prior acts like a \textit{contrarian illusory signal} that causes probability distortions in opposition to the prior and this is a cognitive mechanism that does not require completion of the task. As they gain experience, they keep on applying Bayes rule to update their prior belief both by cues on their current performance and by the prior-dependent contrarian signal.Thus, with the single assumption of intuitive rationality, we can account for all the cognitive biases described on our data within the Bayesian paradigm and integrate the cognitive bias and the learning approach. With this model, and in contrast with \cite{gervais2001}, we don't need to assume a self-attribution bias \citep{langer1975, miller1975} combined with Bayesian learning to produce overconfidence\footnote{Using German survey data about stock market forecasters, \cite{deaves2010} does not confirm that success has a greater impact than failure on self-confidence, which casts doubt on the self-attribution bias explanation.}. Signals of future success and failure are treated symmetrically\footnote{In studies where subjects are free to stay or to leave after a negative feedback, subjects who update most their confidence in their future success to a negative feedback are selectively sorted out of the sample. This creates an asymmetry in measured responses to positive and negative feedback. Such spurious asymmetry does not exist in the present experiment, because subjects who fail to reach one level must drop out of the game.}. Finally, unlike models of confidence management (e.g. \citealt{brunnermeier2005,koszegi2006, mobius2014}), we don't have to postulate that individuals manipulate their beliefs and derive direct utility from optimistic beliefs about themselves.

Section \ref{expe} lays down the structure of the experiment and incentives, and provides the basic descriptive statistics. Our large data set allows a thorough description of confidence biases and a dynamic view of their evolution with experience of the task. Section \ref{biases} describes the confidence biases and learning shown by our data. Four basic facts about confidence are reported from our data: \textit{(i)} limited discrimination among different tasks; \textit{(ii)} miscalibration of subjective probabilities of success elicited by the "hard-easy effect"; \textit{(iii)} differential, ability-dependent, calibration biases known as the Dunning-Kruger (or ability) effect \citep{kruger1999unskilled}; and \textit{(iv)} local, but not global, learning. Section \ref{theory} proposes a new theory of over (under)-confidence among intuitive Bayesians which integrates doubt and learning and can predict biases, before as well as during the task, in repeated as well as in single trials. Doubt-driven miscalibration appears to be a sufficient explanation, not only for the hard-easy effect and the 'ability' or Dunning-Kruger effect, but also for limited discrimination and for the overprecision phenomenon. The theory is further used in section \ref{biases_and_learning} to predict the evolution of confidence over experience on our data set. For instance, low-ability subjects first lose confidence when they discover their low performance during the first and easiest level; but they eventually regain their initial confidence in own ability to perform more difficult tasks in the future after laborious but successful completion of the first level. Intuitive Bayesians exhibit \textit{conservatism}, that is, under-reaction to received information, and slow learning. Finally, we show in sub-section \ref{why} that the cues upon which subjects construct their own estimate of success, \textit{i.e.} confidence, widely differ from the genuine predictors of success, which further explains the \textit{planning fallacy}\footnote{The planning fallacy is the tendency to underestimate the time needed for completion of a task. See, e.g. \citet{buehler2002inside}.}. The conclusion follows in section \ref{conclusion}.

\section{The experiment} \label{expe}
\subsection{Task and treatments} \label{task}

Participants perform a real-effort, rather long and difficult, task for which they get paid according to their degree of success. The task consists in solving anagrams ranked in three levels of increasing difficulty. It is performed during a maximum of 15 rounds lasting no more than 8 minutes each. These 15 rounds are structured in three successive levels of increasing difficulty, designated respectively as the training level, the middle level, and the high level.

Participants are successful at one level when they manage to decode 2/3 of the anagrams at this level. An example of the task screen is reproduced in appendix. The training level consists of 9 rounds of low difficulty (\textit{i.e}. 6 anagrams per round to be solved in no more than eight minutes). It is long enough to let participants feel that a large effort and ability is required of them to succeed at the optional upper levels. It does also let them ample time to learn the task. The middle and high levels, which come next, comprise 3 rounds each.                                                   

The gradient of task difficulty was manipulated after completion of the training level and two conditions are available: \textit{(i)} in the \textit{'wall'} condition, the difficulty jumps sharply at middle level, but remains constant at high level; \textit{(ii)} in the \textit{'hill'} condition, the difficulty always rises from one level to the next, slowly first at middle level, then sharply at high level.

\begin{figure}[h!]
\begin{center}
\caption{Decision problem perceived by participants at the start of level 2 of the choice treatment.}
\label{choice}
\begin{minipage}{1\textwidth} % choose width suitably
\includegraphics[width=\linewidth]{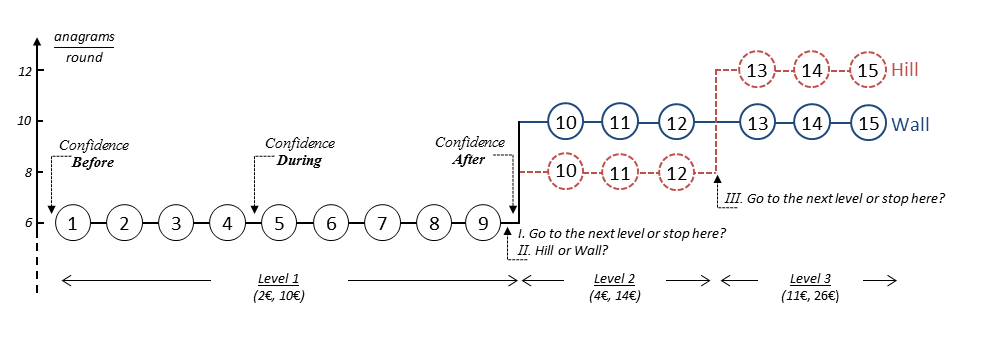}
{\scriptsize Notes:\\Payoffs in parentheses : (fail, success and stop).\\
Decisions I, II and III  are conditioned to success in the previous level.\\
Decision II depends on the treatment.\\
Estimation of Confidence After is conditioned to success in the first and decision to start the second level.
\par}
\medskip 
\end{minipage}
\end{center}
\end{figure}

By the end of the experiment, the required number of anagrams is the same for the \textit{'wall'} and \textit{'hill'} conditions. However, the distribution of anagrams to be decoded differs for these two conditions. In the wall condition, ten anagrams per round are proposed at the middle and high levels, of which 20 anagrams at least must be decoded per level. In the hill condition, eight anagrams per round are proposed at middle level, and this rises to twelve anagrams at high level. Decoding sixteen anagrams in three rounds is required for middle level; and decoding twenty-four anagrams in three rounds is required for high level. This design can be visualized in Figure \ref{choice}. The same figure appears (without the legends) on the screen before each round\footnote{The screen highlights the round, the number of correct anagrams cumulated during the current level and the number of anagrams needed to pass this level.}.

The manipulation of the 'wall' and 'hill' conditions gave rise to three treatments:
\begin{itemize}
\item Wall treatment (\textit{wall}): the wall condition is imposed to participants who passed the training level;
\item Hill treatment (\textit{hill}): the hill condition is imposed to participants who passed the training level;
\item Choice treatment (\textit{choice}): a choice among the two conditions (\textit{wall} or \textit{hill}) is proposed to participants who passed the training level.
\end{itemize}

The double or quits game is played under these three treatments. All subjects first go through the training level. Those who were successful -i.e., those who solved at least 36 anagrams during the training level- will then be asked to double or quits:

\begin{itemize}
\item \textbf{Double}: Continue to the next level to win a substantial increase in earnings; 
\item \textbf{Quits}: Stop the experiment and take your earnings.  
\end{itemize}
	
Participants who decide to go to middle level get a consolation prize that is lower than the foregone earnings if they fail or drop out before the third round. If they succeed middle level, they will be asked again to double or quits. The same rules apply for high level at rising levels of earnings. The potential gains (in Euros) were (10, 2) at the training level, that is, 10\euro \ for successful quitters and 2\euro \ for failures, (14, 4) at middle level, and (26, 11) at high level. 

\subsection{Experimental sessions} \label{sessions}

We ran 24 sessions for a total of 410 participants, half for the choice treatment and the other half equally split between the \textit{'wall'} and \textit{'hill'} treatments. Eight sessions were run in the BULCIRANO lab (Center for Interuniversity Research and Analysis on Organizations), Montreal (Canada), and the same number of sessions were conducted at the LEEP (\textit{Laboratoire d'Economie Exp\'erimentale de Paris}), Pantheon-Sorbonne University. The difference between Paris and Montreal was observed to be insignificant. Thus, eight additional sessions were conducted at LEEP in order to acquire robust results. A show-up fee of 5\euro \ in Paris and Can\$ 5 in Montreal was paid to the participants (from now on, all money amounts will be given in Euros). About 80\% of the participants were students. 

At the start, instructions were read out and a hard copy of it was also provided individually. Participants answered six questions to test their full comprehension of the experiment. Information on gender, age, educational level and labor market status was required. The last question was a hypothetical choice between 5\euro \ for sure and an ambiguous urn containing 100 balls of two colors (white and black) in unknown proportions. Ten Euros (10\euro) were to be earned if a black ball was drawn. Choice of the sure gain provided a rough but simple measure of risk aversion in the uncertainty context of the experiment.

\subsection{Descriptive statistics} \label{stats}

The main descriptive statistics for the three treatments are reported in Table \ref{statdes}: 

\begin{table}[!ht]
\begin{center}
\caption{\label{statdes} \small{Descriptive statistics for the three treatments}}
% Table generated by Excel2LaTeX from sheet 'table1'
\footnotesize
\begin{tabular}{rrrr}
\hline
           & \multicolumn{ 3}{c}{{\bf Treatments}} \\

{\bf Variables} & {\bf Wall} & {\bf Hill} & {\bf Choice} \\
\hline
      Male &       56\% &       48\% &       49\% \\

       Age &       24.5 &       25.8 &       25.1 \\

Risk Averse &       54\% &       59\% &       51\% \\

  Payments &        9.1 &        8.9 &        7.8 \\

Total anagrams solved &       55.6 &       53.7 &       54.3 \\
Ability\tablefootnote{Ability is measured by the number of anagrams solved per minute in the first 4 rounds. It lies in the interval [0,6].} &        2.8 &        2.7 &        2.6 \\

{\bf Number of observations} &  {\bf 101} &  {\bf 106} &  {\bf 203} \\
\hline
\multicolumn{ 4}{l}{{\it Decision to double conditional on success at previous level:}} \\
\hline
Middle level &  78\% (91) &  76\% (90) & 77\% (176) \\

High level &  95\% (22) &  72\% (29) &  82\% (34) \\
\hline
\multicolumn{4}{l}{
\begin{minipage}{0.72\textwidth}%
\medskip
\scriptsize Notes: Decision to double to High level: difference between the "Wall" and "Hill" treatments is significant at 5\%; all other differences are not significant at 10\% level (t-test). Number of participants successfully clearing the previous level  is in parentheses.
\end{minipage}
  } %}\\
\end{tabular}
\end{center}
\end{table}

%\normalsize
%\end{tabular}  
%\normalsize
%\end{center}
%\footnotesize {} 

The results of tests show that the three samples are homogeneous. No significant difference is observed among the samples' means for individual characteristics. As expected, the \textit{'wall}' and \textit{'hill'} treatments had a substantial impact on the decision to double upon reaching the middle level. Almost everybody doubles in the \textit{'wall'} treatment on reaching middle level because the high level is no more difficult than the middle level. In contrast, only 72\% enter the high level in the \textit{'hill'} treatment as the difficulty gradient is very steep  (t-test: \textit{t}= 2.20; \textit{p-value}=0.033). In spite of these differences, the number of anagrams solved and payments may be considered equal among treatments at the usual level of significance.

Subjects can also be grouped in three different levels of ability, according to the number of anagrams solved per minute in the first 4 rounds: high ability (first tercile), medium ability (second tercile) and low ability (last tercile). Some descriptive statistics for the three treatments are reported on Table \ref{statability}. The three groups are homogeneous in terms of gender and risk aversion but a slightly greater proportion of low-ability subjects can be found among older, probably non-student, participants.

\begin{table}[!ht]
\begin{center}
\caption{\label{statability} \small{Descriptive statistics by ability level}}

% Table generated by Excel2LaTeX from sheet 'table1'
\footnotesize

\begin{tabular}{rrrrrrr}
\hline
           & \multicolumn{ 3}{c}{{\bf Level of ability}} & \multicolumn{ 3}{c}{{\it Difference}} \\

{\bf Variables} & {\bf High} & {\bf Medium} &  {\bf Low} &  {\it M-H} &  {\it L-M} &  {\it L-H} \\
\hline
      Male &       47\% &       54\% &       50\% &   {\it ns} &   {\it ns} &   {\it ns} \\

       Age &       23.6 &       24.5 &       27.2 &   {\it ns} &  {\it ***} &  {\it ***} \\

Risk Aversion &       53\% &       50\% &       59\% &   {\it ns} &   {\it ns} &   {\it ns} \\

  Payments &       11.7 &        7.7 &        6.0 &  {\it ***} &   {\it **} &  {\it ***} \\

Number anagrams solved &       67.7 &       53.8 &       42.6 &  {\it ***} &  {\it ***} &  {\it ***} \\

   Ability &        4.5 &        2.4 &        1.1 &  {\it ***} &  {\it ***} &  {\it ***} \\

{\bf Number of observations} &  {\bf 131} &  {\bf 142} &  {\bf 137} &            &            &            \\
\hline
 \multicolumn{ 7}{l}{{\it Decision to double conditional on success at previous level:}} \\
\hline
Middle level & 91\% (128) & 81\% (127) & 54\% (102) &   {\it **} &  {\it ***} &  {\it ***} \\

High level &  87\% (55) &  72\% (25) &   80\% (5) &   {\it *} &   {\it ns} &   {\it ns} \\
\hline
\multicolumn{7}{l}{
\begin{minipage}{1\textwidth}%
\medskip
\scriptsize Notes: Significance level: * 10\%; ** 5\%; ***1\%; ns: not significant at 10\% level (t-test). Number of participants successfully clearing the previous level  is in parentheses
\end{minipage}
  } %}\\
\end{tabular}
\end{center}
\end{table}

Table \ref{statability} shows that "ability" strongly discriminates among participants in terms of performance (total anagrams solved, payments) and quits before the middle level. However, the training level was meant to be easy enough that three-quarters (102:137) of low-ability subjects would pass it.

\subsection{Confidence judgments} \label{judgments}

Participants were asked to state their subjective probability of success for the three levels and at three moments: before, during, and after the training level. Before beginning the game, they were shown a demonstration slide which lasted one minute. Anagrams of the kind they would have to solve appeared on the screen with their solution. Then, they were asked to assess their chances of success on a scale of 0 to 100 \citep{adams1957}, and the game started for real. After four rounds of decoding anagrams, players were asked again to rate their confidence. Lastly, players who had passed the training level and decided to double re-estimated their chances of success for the middle and high levels.

The Adams's (\citeyear{adams1957}) scale that we used is convenient for quantitative analysis because it converts confidence into (almost) continuous subjective probabilities. It was required for consistency that the reported chances of success do not increase as the difficulty level increased. Answers could not be validated as long as they remained inconsistent. Subjects actually used the whole scale but, before the experiment, 14\% expressed absolute certainty that they would succeed the first level and only 1 participant was sure that she would fail.

We did not directly incentivize beliefs because our primary aim was not to force subjects to make optimal forecasts of their chances of success but to have them report sincerely their true beliefs in their attempt to maximize their subjective expected utility, and to observe the variation of such beliefs with experience. The true beliefs are those which dictate actual behavior following such prediction, and the latter was incentivized by the money gains based on subjects' decisions to double or quits and performance in the task. \cite{armantier2013} have recently generalized previous work on proper scoring rules (see their extensive bibliography). They show that, when subjects have a financial stake in the events they are predicting and can hedge their predictions by taking additional action after reporting their beliefs, use of any proper scoring rule generates complex distortions in the predictions and further behavior since these are not independent and are in general different from what they would have been if each had been decided separately. In the present context, final performance yields income and does not immediately follow the forecast. Hence, incentivizing forecasts might force subjects to try and adjust gradually their behavior to their forecast and, therefore, unduly condition their behavior. A further difficulty encountered in this experiment was that, by incentivizing beliefs on three successive occasions, we induced risk-averse subjects to diversify their reported estimates as a hedge against the risk of prediction error. Self-report methods have been widely used and validated by psychologists and neuroscientists; and recent careful comparisons of this method with the quadratic scoring rule\footnote{After the subject has reported a probability $ p $, the quadratic scoring rule imposes a cost that is proportional to $ (1-p)^{2} $ in case of success and to $ (0-p)^{2} $ in case of failure. The score takes the general form: $ S= a-b $. Cost, with $ a, b>0 $. } found that it performed as well \citep{clark2009} or better \citep{hollard2015} than the quadratic scoring rule\footnote{The second study also included the lottery rule in the comparison and found that the latter slightly outperformed self-report. The lottery rule rests on the following mechanism: after the subject has reported a probability \textit{p}, a random number \textit{q} is drawn. If \textit{q} is smaller than \textit{p}, the subject is paid according to the task. If q is greater than \textit{p}, the subject is paid according to a risky bet that provides the same reward with probability \textit{q}. The lottery rule cannot be implemented on our design. }. Considering that self-reports perform nicely while being much simpler and faster than incentive-compatible rules, use of the self-report seemed appropriate in this experiment.   

\section{Describing confidence biases and learning} \label{biases}

\subsection{Limited Discrimination}
About half of our subjects were selected randomly into the 'wall' and 'hill' treatments and could not choose between the two. Those selected in one path were informed of the characteristics of their own path but had no knowledge whatsoever of the characteristics, nor even the existence, of the other path.
\\
\paragraph{Result 1 (Limited discrimination):} \textit{Subjects do not perceive differences of difficulty between two different tasks in the future unless such differences are particularly salient. Moreover, they are not forward-looking, in the sense that they are unable to anticipate the increased likelihood of their success at the high level conditional on passing the middle level. However, they can be sophisticated when it is time for them to choose.}

\paragraph{Support of result 1:} Table \ref{conf_compare} compares confidence judgments regarding the three levels of difficulty among the '\textit{wall}' and the '\textit{hill}' subjects before, during, and after the training period. Although the '\textit{wall}' and '\textit{hill}' were designed to be quite different at the middle and high levels, the subjective estimates of success exhibit almost no significant difference at any level. The single exception concerns the early estimate (before round 1) regarding the high level for which the difference of gradient between the two paths is particularly salient. However, the difference ceases to be significant as subjects acquire experience of the task. This striking observation suggests that individuals are unable to discriminate distinctive characteristics of the task unless the latter are particularly salient.

Perhaps even more disturbing is the fact that, in Table \ref{conf_compare}, subjects discount their confidence level from the middle to the high level as much in the Wall as in the Hill treatment. For instance, just before the middle level, the ratio of confidence in passing the high level to confidence in passing the middle level was close to 0.70 in both treatments. However, a perfectly rational agent should realize that the high level is no more difficult than the middle level in the Wall treatment whereas it is much more difficult in the Hill treatment. Thus, she should report almost the same confidence at both levels in the Wall treatment, and a considerably lower confidence at the high level in the Hill treatment. The latter observation suggests that most individuals are unable to compute conditional probabilities accurately even when the latter is equal to one as in the Wall treatment. They don't anticipate that, if they demonstrate the ability to solve 20 anagrams or more at middle level, they should be almost sure to solve 20 or more at the high level. 
However, subjects do make the right inference when it is time for them to make the decision since 95\% of subjects who passed the middle level in the Wall treatment decided to continue (Table \ref{statdes}). And, if they have a choice between Wall and Hill, they do make a difference between these two tracks: 71.4\% of doublers then prefer the Wall track although they would have greater chances of success at the middle level if they chose Hill. This observation suggests that subjects did not maximize their immediate probability of success but made a sophisticated comparison of the expected utility of both tracks, taking the option value of Wall in consideration before making an irreversible choice of track spanning over two periods\footnote{We are grateful to Luis Santos-Pinto for making the last point clear in early discussions.}.

\begin{table}[!ht]
\begin{center}
\caption{\label{conf_compare} \small{A comparison of confidence for the wall and hill treatments shown separately}}

% Table generated by Excel2LaTeX from sheet 'table1'
\footnotesize
\begin{tabular}{lrccr}
\hline
{\it {\bf Subjective confidence}} &            & \multicolumn{ 2}{c}{{\bf No-choice treatment}} &            \\

           &            &  Wall (\%) &  Hill (\%) & {\it Difference} \\
\hline
\hline
{\it {\bf Before round 1:}} & {\it Level 1} &         80 &         77 &   {\it ns} \\

           & {\it Level 2} &         62 &         58 &   {\it ns} \\

           & {\it Level 3} &         47 &         40 &   {\it **} \\
\hline
{\it {\bf Before round 5:}} & {\it Level 1} &         71 &         71 &   {\it ns} \\

           & {\it Level 2} &         53 &         52 &   {\it ns} \\

           & {\it Level 3} &         40 &         36 &   {\it ns} \\
\hline
{\it {\bf Before round 10:}} & {\it Level 2} &         60 &         56 &   {\it ns} \\

           & {\it Level 3} &         43 &         39 &   {\it ns} \\
\hline
\multicolumn{5}{l}{
\begin{minipage}{0.8\textwidth}%
\medskip
\scriptsize Notes. Observations: Before rounds 1 and 5 (before round 10): 101 (71) for wall and 106 (68) for hill. \textbf{Significance Level}: * \(p<.10\), ** \(p<.05\), *** \(p<.01\), ns: not significant at 10\% level.
\end{minipage}
  } %}\\
\end{tabular}
\end{center}
\end{table}

\subsection{Miscalibration}

\begin{figure}[!t]
\begin{center}
\caption{Hard-easy effect observed at three levels}
\label{hardeasy}
\begin{minipage}{0.8\textwidth} % choose width suitably
\includegraphics[width=\linewidth]{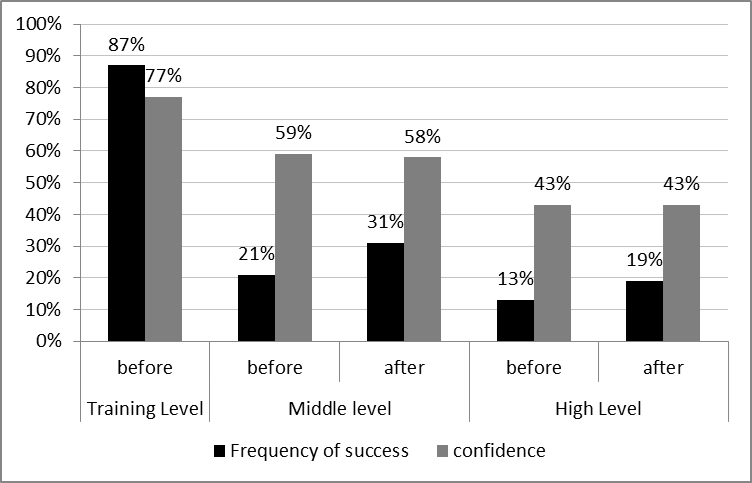}
{\scriptsize Notes: \textbf{Observations}: before training level (N: 410); after training level (N: 275 - analysis restricted to doublers). \textbf{Differences} between frequency of success and confidence (before and after) are significant at 1\% at all levels (Training, Middle and High). (\textit{t-test})
\par}
\medskip 
\end{minipage}
\end{center}
\end{figure}

\paragraph{Result 2 (The hard-easy effect):} \textit{In comparison with actual performance, confidence in one's ability to reach a given level is underestimated for a novel but relatively easy task (the training level); and it is overestimated for the subsequent more difficult tasks (the middle and high levels). Overconfidence increases in relative terms with the difficulty of the task.
}
\textit{Conditional on an initial success (training level) and on the decision to continue, confidence in one's ability to reach higher levels is still overestimated. Thus, initially successful subjects remain too optimistic about their future. }
\\
\paragraph{Support for result 2:} Figure \ref{hardeasy} compares the measured frequency of success with the reported subjective confidence in the three successive levels of increasing difficulty. For the middle and high levels, we also indicate these two probabilities as they appear before the training period and after it conditional on doubling. The Choice and No-choice conditions have been aggregated on this figure because no significant difference was found in the result of tests.

The task required at the training level was relatively easy for our subjects since 87\% passed this level. However, subjects started it without knowing what it would be like and, even after four rounds of training, they underestimated their own ability to a low 77\% probability of success. The difference among the two percentages is significant (t=5.77, p=0.000; t-test). Hence, individuals are under-confident on the novel but relatively easy task.

In contrast, subjects appear to be overconfident as the task gets increasingly difficult. They consistently diminish their estimated probabilities of success but do not adjust their estimates in proportion to the difficulty of the task. Thus, individuals tend to overestimate their own chances for the advanced levels. The difference between the frequency of success and confidence before the task is always significant, both at the middle level (\textit{t=18.3, p=0.000}) and at the high level (\textit{t=17.1, p=0.000}).

The same conclusions hold conditional on passing the training level and choosing to double. Subjects remain overconfident in their future chances of success. However, their confidence does not rise after their initial success in proportion to their chances of further success.

\subsection{The ability effect}
\paragraph{Result 3 (The ability effect):} \textit{Overcalibration diminishes with task-specific ability.}

\paragraph{Support for result 3:} The hard-easy effect is reproduced on Figures \ref{he1}, \ref{he2}, \ref{he3} for the three ability terciles\footnote{Difference between confidence and frequency of success is significant at 1\% for all ability levels. For these figures, we selected confidence reported after $4^{th}$ round (during training level) in order to minimize the impact of mismeasurement.}. Low-ability subjects are obviously more overconfident at middle and high levels relative to high and medium-ability individuals. This result confirms earlier observations of \cite{kruger1999unskilled} among others (see \cite{ryvkin2012} for a recent overview and incentivized experiments). The so-called Dunning-Kruger effect has been attributed to a metacognitive inability of the unskilled to recognize their mistakes\footnote{\label{realeffort}The Dunning-Kruger effect initially addressed general knowledge questions whereas we consider self-assessments of own performance in a real-effort task.}. We give here another, and in our opinion, simpler explanation\footnote{Our explanation may also be better than the initial explanation such that the unskilled are unaware of their lower abilities. \cite{miller2011} found that students with poor abilities showed greater overconfidence than high-performing students, but they also reported lower confidence in these predictions.}. The ability (or Dunning-Kruger) effect may be seen as a corollary of the hard-easy effect because "difficulty" is a relative notion and a task that a low-ability individual finds difficult certainly looks easier to a high-ability person. Thus, if overconfidence rises with the difficulty of a task, it is natural to observe that it declines on a given task with the ability of performers.

%\FloatBarrier

\begin{subfigures}

\begin{figure}[!t]\centering
\caption{Under-confidence at the training level, by ability. }\label{he1}
\includegraphics[scale=0.7]{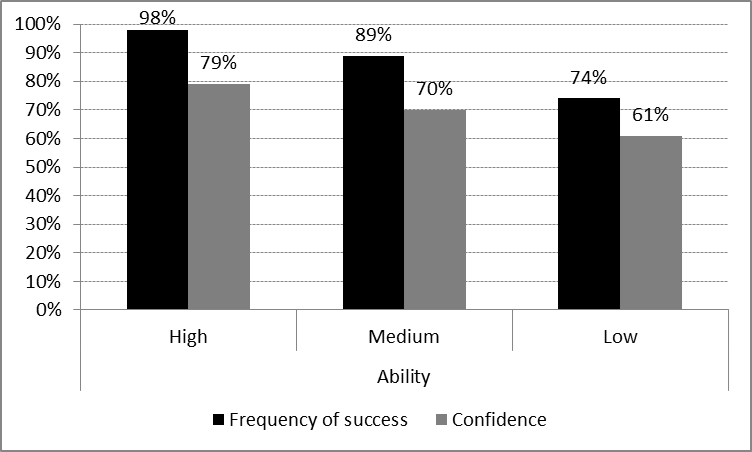}
\end{figure}
\begin{figure}\centering
\caption{Overconfidence at middle level, by ability.   }\label{he2}
\includegraphics[scale=0.7]{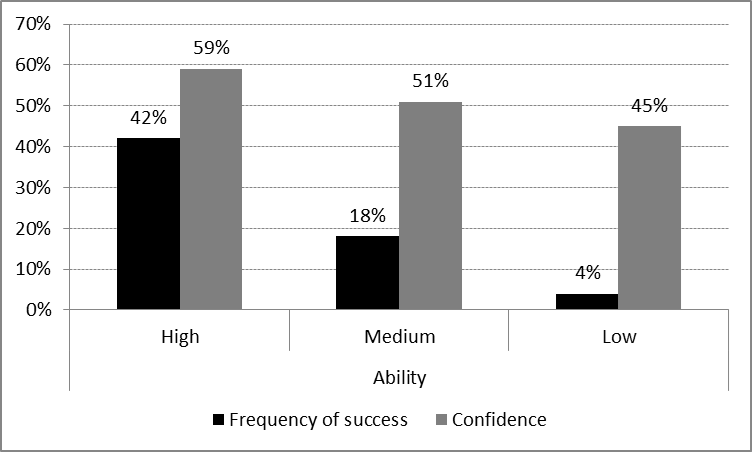}
\end{figure}
\begin{figure}\centering
\caption{Overconfidence at high level, by ability.  }\label{he3}
\includegraphics[scale=0.7]{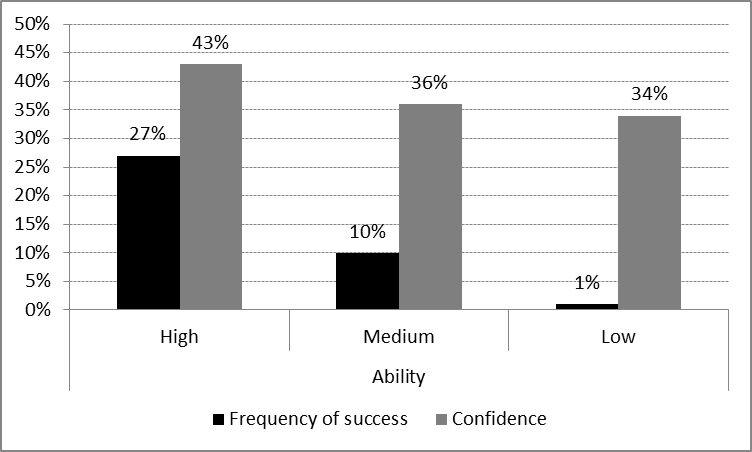}
\end{figure}
\end{subfigures}

\FloatBarrier

\subsection{Learning}\label{Learning}

\paragraph{Result 4 (Learning is local, not global):} \textit{Confidence and performance co-vary during the task. Subjects learned \textbf{locally} upon experiencing variations in their performance. However, they didn't learn \textbf{globally} in our experiment, since doublers remained as confident as before after completing the training level irrespective of their true ability level.}

\begin{figure}[!h]
\begin{center}
\caption{Variation of confidence with experience, by level of ability: middle level}
\label{middle}
\begin{minipage}{0.8\textwidth} % choose width suitably
\includegraphics[width=\linewidth]{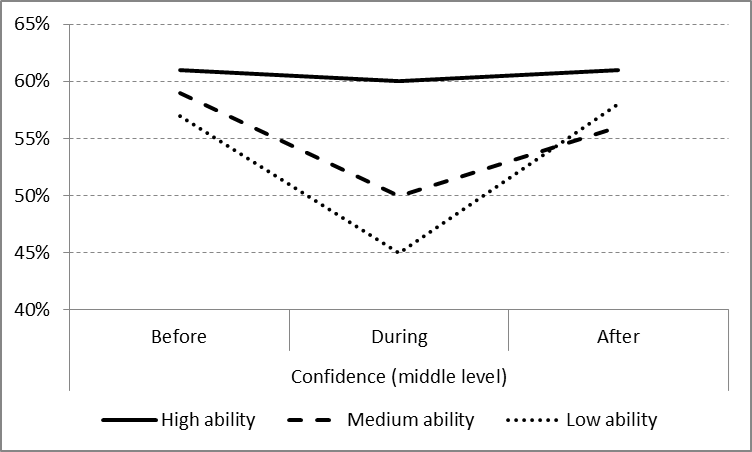}
{\scriptsize Notes. \textbf{Sample size}: 410 individuals for Before and During, and 275 for After (only doublers). We report the adjusted ability for doublers, see Footnote \ref{adjust} for more details. \textbf{Differences between ability levels} are significant at 1\% level Before and During. Differences After are not significant at 10\% level. \textbf{Differences by ability level}: High-ability: During-Before: ***; After-During: ns; After-Before: ns. Medium-ability: During-Before: ***; After-During: ***; After-Before:**. Low- ability: During-Before:***; After-During: *** ; After-Before: ns. \textbf{Significance level}: *** 1\%; ** 5\%; * 10\%; ns:  not significant at 10\% level (t-test).
\par}
\medskip 
\end{minipage}
\end{center}
\end{figure}

\begin{figure}[!h]
\begin{center}
\caption{Variation of confidence with experience, by level of ability: high level}
\label{high}
\begin{minipage}{0.8\textwidth} % choose width suitably
\includegraphics[width=\linewidth]{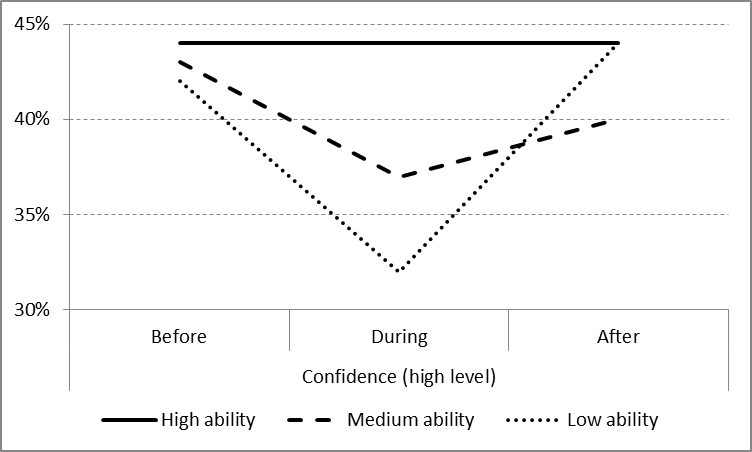}
{\scriptsize Notes. \textbf{Sample size}: 410 individuals for Before and During, and 275 for After (only doublers). We report the adjusted ability for doublers, see Footnote \ref{adjust} for more details. \textbf{Differences between ability levels} are significant at 1\% level Before and During. Differences After are not significant at 10\% level. \textbf{Differences by ability level}: High-ability: During-Before: ns; After-During: ns; After-Before: ns. Medium-ability: During-Before: ***; After-During: **; After-Before: ns. Low-ability: During-Before:***; After-During: *** ; After-Before: ns.. \textbf{Significance level}: *** 1\%; ** 5\%; * 10\%; ns:  not significant at 10\% level (t-test).
\par}
\medskip 
\end{minipage}
\end{center}
\end{figure}

\begin{figure}[!h]
\begin{center}
\caption{Number of anagrams solved per round by level of ability}
\includegraphics[scale=0.8]{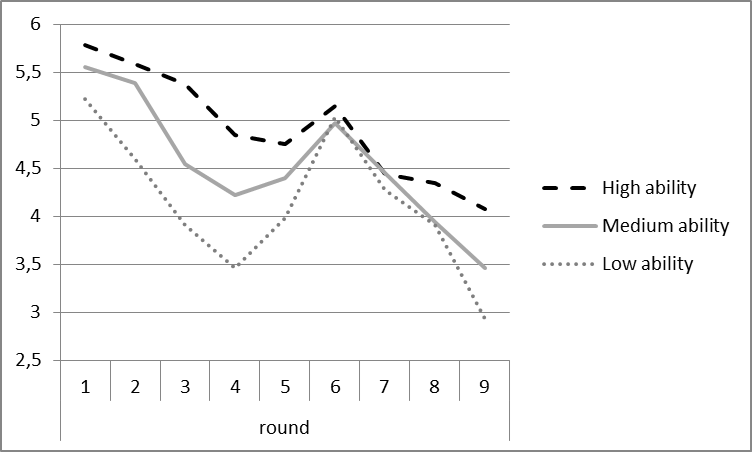}
\label{effort}
\end{center}
\end{figure}

\noindent
\paragraph{Support for result 4:} Figures \ref{middle} and \ref{high} describe confidence by ability group before, during, and after the training period\footnote{No significant difference was found between the Choice and No-choice conditions, suggesting that the option to choose the preferred path does not trigger an illusion of control. }\footnote{\label{adjust} Participants who reported confidence after the training period were more able than average since they had passed this level and decided to double. Thus, we compare ability-adjusted confidence Before and During with the reported confidence After. The ability-adjusted confidence Before and During are obtained by running a simple linear regression of confidence Before and During on ability, measured by the average number of anagrams solved per minute in the first 4 rounds of the training level. The estimated effect of superior ability of doublers was added to confidence During or Before to get the ability-adjusted confidence which directly compares with the observed confidence After. } for the middle and high level respectively whereas Figure \ref{effort} describes the variation of performance of the same groups within the same period. These graphs, taken together, show a decline in both (ability-adjusted) confidence and performance during the first four rounds, followed by a concomitant rise of confidence and performance in the following rounds\footnote{With a single exception, confidence variations are statistically significant at 1\% level in the middle and high levels.}. The observed decline of confidence at the beginning of the training period can be related on Figure \ref{effort} to the fact that participants solved less and less anagrams per period during the first four periods: 5.51 on average in period 1, 5.18 in period 2, 4.60 in period 3, and 4.17 in period 4\footnote{There was no significant difference between treatments.}. Subjects kept solving at least two-thirds of the anagrams available during the training session but probably lost part of their motivation on repeating the task. On sequentially observing their declining performance, they revised their initial estimate of future success downward. However, on being asked to report their confidence after four rounds, they became conscious of their performance decline and responded to this information feedback. Performance rose sharply but momentarily during the next two rounds. The average performance first rose to 4.37 in period 5 and 5.05 in period 6 then sharply declined to 4.39 in period 7, 4.06 in period 8 and 3.48 in period 9. As soon as subjects became (almost) sure of passing the training level, they diminished their effort. During the experiment it was also observed that individuals stopped decoding further anagrams as soon as the minimum requirement to clear a level was fulfilled. 

Subjects experiencing low (medium) performance in the first rounds seem to learn locally that they have a low (medium) ability since the confidence gap widens during the first four periods. However, this learning effect is short-lived since the confidence gap shrinks back to its initial size after low (medium)-ability subjects strove to succeed, increasing their performance (as reported on Figure \ref{effort}) and regaining confidence. Eventually, experienced "doublers" are as confident to succeed at higher levels as they were before the task, irrespective of their ability level: there is no global learning effect. We share the conclusion of \cite{merkle2011} that the persistence of prior beliefs is inconsistent with fully rational-Bayesian behavior(see also \citealt{benoit2015}).

\FloatBarrier

\section{Theory} \label{theory}

We present now a simple Bayesian model that describes absolute confidence reported before and during completion of a task, and predicts limited discrimination, the hard-easy effect and the ability effect. It builds on ideas put forward by \citet{erev1994} and \citet{moore2008} who both consider that confidence, like most judgments, are subject to errors. \citet{erev1994} view confidence as a subjective probability that must lie between 0 and 1. Hence, probabilities close to 1 are most likely to be underestimated and probabilities close to 0 are most likely to be overestimated. The hard-easy effect and the ability effect may be merely the consequence of that simple truth. However, their theory offers a qualitative assessment that lacks precision and cannot be applied to intermediate values of confidence. \citet{moore2008} analyze confidence as a score in a quiz that the player must guess \textit{after} completion of the task and \textit{before} knowing her true performance. Bayesian players adjust their prior estimate after receiving a subjective signal from their own experience. It is natural to think that signals are randomly distributed around their true unknown value. Assuming normal distributions for the signal and the prior, the posterior expectation of confidence is then a weighted average of the prior and the signal lying necessarily between these two values. Thus, if the task was easier than expected, the signal tends to be higher than the prior. The attraction of the prior pulls reported confidence below the high signal, hence below true performance on average since the signal is drawn from an unbiased distribution. While rational-Bayesian models like \citet{moore2008} may account for learning over experience, they fail to predict limited discrimination, miscalibration of confidence before completion of the task, or the absence of global learning. Therefore, we add to the Bayesian model a crucial but hidden aspect of behavior under risk or uncertainty, that is doubt. We describe the behavior of subjects who are uncertain of their true probability of success and become consequently vulnerable to prediction errors and cognitive illusions if they rely essentially on what they perceive sequentially. We designate these subjects as "intuitive Bayesians". It turns out, unexpectedly, that the same model also predicts the overprecision bias of confidence, which we consider as a further confirmation of its validity.

Intuitive Bayesians may miscalibrate their own probability of success even if they have an unbiased estimate of their own ability to succeed. This can occur if they are uncertain of the true probability of success because they can be misled by "available" illusory signals triggered by their doubt. The direction of doubt is entirely different depending on whether their prior estimate led them to believe that they would fail or that they would succeed. We thus distinguish miscalibration among those individuals who should normally believe that they should not perform the task and those who should normally believe that they should.

To facilitate intuition, let us first consider a subject who is almost sure to succeed a task, either because the task is easy or because the subject has high-ability (\textit{H}). However, the "availability" of a possible failure acts like a negative signal which leads to overweighting this possibility \citep{tversky1973}, and underweighting her subjective probability of success\footnote{  The time $t=(1,5,10)$ when confidence is reported is omitted in this sub-section to alleviate notations.  } $Ep_H$, \textit{i.e.} underconfidence:  

\begin{center}
\begin{equation} \label{eq1}
q_{H}=\mu Ep_H+(1-\mu)0= \mu Ep_H\leq Ep_H, 
\end{equation}
with $0<\mu\leq 1$
\end{center}

Even though high-ability agents are almost sure of succeeding the training level, their confidence is way below 1, confirming the Dunning-Kruger effect where high-ability subjects underestimate their abilities. An estimate of this undercalibration bias for an easy task is derived from Figure \ref{he1}:

\begin{center}
$ \mu_{H_{ \textrm{\textit{(training level)}}}}= \dfrac {0.79}{0.98}=0.806\cong q_{H_{ \textrm{\textit{(training level)}}}}$ 
\end{center}

The undercalibration bias is: $1-0.806=0.194$. 

However, underweighting a high probability of success need not reverse the intention of doubling. Indeed, taking the expected value as the decision criterion, among 167 "able" subjects who should double if objective probabilities are used for computation, 158 (i.e. 94.6\%) still intended to double according to the subjective confidence reported before the game\footnote{Very close numbers are obtained for all calibration biases with confidence reported during the game.}.

At the other end of the spectrum, consider now a subject who is almost sure of failing, either because the task is very difficult or because the subject has low-ability (\textit{L}). However, the "availability" of a possible success leads to overweighting her subjective probability of success $Ep_L$ \textit{i.e.} overconfidence:

\begin{center}
\begin{equation} \label{eq2}
q_{L}=\mu Ep_L+(1-\mu)1\geq Ep_L, 
\end{equation}
with $0<\mu \leq 1$
\end{center}
                                       Thus, even though low ability agents should give up a difficult task, they are overconfident and are thus tempted by the returns to success\footnote{This should not be confounded with motivated inference as it applies symmetrically to undesirable and desirable outcomes. }. In the limit, confidence remains positive if one is almost certain to fail. This means that low-ability individuals always exhibit a positive bottom confidence, which is in line with the Dunning-Kruger effect (they overestimate their abilities). An estimate of this overcalibration bias for the high level is derived from Figure \ref{he3}: 

\begin{center}
$ 1-\mu_{L_{ \textrm{\textit{(high level)}}}}= \dfrac{0.34-0.01}{1-0.01}=0.333\cong q_{L_{ \textrm{\textit{(high level)}}}}$ 
\end{center}
Similarly, the overcalibration bias for the middle level is derived from Figure \ref{he2}: 

\begin{center}
$ 1-\mu_{L_{ \textrm{\textit{(middle level)}}}}'= \dfrac{0.45-0.04}{1-0.04}=0.427\cong q_{L_{ \textrm{\textit{(middle level)}}}}$  
\end{center}

Notice that the overcalibration bias is about twice as large as the undercalibration bias. Hence, taking the expected value as the decision criterion, among 190 "unable" subjects who should quit if objective probabilities are used for computation, 159 (i.e. 83.7\%)  intended to double according to the subjective confidence reported before the game.   

To sum up, we explain both the hard-easy effect and the ability effect by an availability bias triggered by the doubt about one's possibility to fail a relatively easy task (underconfidence) or to succeed a relatively difficult task (overconfidence). If probabilities are updated in a Bayesian fashion, the calibration bias is the relative precision of the illusory signal. The latter is inversely related with the absolute precision of the prior estimate and positively related with the absolute precision of the illusory signal. Thus, we mustn't be surprised to find that our estimate of the calibration bias is lower for the training level (19.4\%) than for upper levels (42.7\% and 33.3\% respectively) because experience in the first rounds of the training level must be more relevant for predicting the probability of success in the training level than in subsequent levels. And, when comparing upper levels, the illusion of success should be more credible for the near future (middle level) than for the more distant future (high level).

This explanation is also consistent with the other measures displayed by Figures \ref{he1}, \ref{he2}, \ref{he3}, given the fact that they aggregate overconfident subjects who should not undertake the task with underconfident subjects who should undertake it\footnote{{The rational decision to undertake a non-trivial task of level \textit{l} (with a possibility to fail and regret) is subjective. The economic criterion for making this decision rests on the comparison of the expected utilities of all options conditional on the estimated probabilities of success at the time of decision. A rational subject should refuse the task if the expected utility of continuing to level \textit{l} or above is no higher than the expected utility of stopping before level l. We make use of this criterion for writing equations \ref{eq6} and \ref{eq7} in the next sub-section (\ref{updating}).}\label{rationaldecision}}. If $\lambda _{L}$ is the proportion who should stop and $\lambda _{H}$ the proportion who should continue ($\lambda _{L}+\lambda _{H}\equiv 1$), the average confidence is: $\lambda _{L}(\mu Ep_{L}+1-\mu )+\lambda _{H}\mu Ep_{H}=\mu \overline{Ep}+(1-\mu )\lambda _{L}$. Confidence is overcalibrated on average iff $\lambda _{L}>\overline{Ep}$ and undercalibrated iff the reverse condition holds. The apparent overcalibration of confidence for a difficult task takes less extreme values when the average measured ability of the group rises. For instance, the results displayed by Figure \ref{he3} are consistent with our estimate for the overcalibration bias if the proportion of successful middle-ability subjects is 12\% and that of successful high-ability subjects is 25\%, since these two predicted values are close to the observed frequency of success in these groups, respectively 10\% and 27\%.

Remarkably, this simple model of miscalibration also predicts limited discrimination. Although Wall is more difficult than Hill at the middle level, our subjects attributed on average about the same confidence level to both tasks (see table \ref{statdes}). High-ability subjects who should double at middle level in the Wall condition, and low-ability subjects who should stop  before middle level in the Hill condition would both estimate their chances of success  to be higher with 16 anagrams to solve with Hill than with 20 anagrams with Wall. The former would underestimate their chances according to (\ref{eq1}) and the latter would overestimate them according to (\ref{eq2}), but the difference between the two estimates would be the same, equal to $\mu(E_{p_{Hill}}-E_{p_{Wall}}$). Thus, if their prior estimates were unbiased, intuitive (s.t. $\mu<1$) high and low-ability subjects would imperfectly discriminate between Hill and Wall by underestimating the difficulty gap between them.  Things are even worse for middle-ability subjects who should opt for middle level under Hill and quit before middle level under Wall. According to (\ref{eq1}) and (\ref{eq2}), those individuals would have a low estimate ($\mu E_{p_{Hill}}$) of their pass rate under Hill and a high estimate ($E_{p_{Wall}} + 1-\mu$) under Wall. They would then underestimate the difficulty gap more severely than high or low-ability subjects and they might even give a higher estimate under Wall than under Hill\footnote{It is assumed here, as in Table \ref{statdes}, that the two estimates are independent.} iff $E_{p_{Hill}}-E_{p_{Wall}} <((1-\mu)/\mu$). Therefore, our model implies limited discrimination of differences in difficulty by intuitive Bayesians when the difference is not very salient.

A further implication of Bayesian updating is that, in the subject's mind, the precision of the posterior estimate for probabilities of success, \textit{i.e.} confidence in her estimate, is increased by reception of the illusory signal, whatever the latter may be\footnote{If $\nu _{i}$ denotes the prior precision of subject $I's$ estimate of her future success (omitting level $l$ for simplicity) $\nu _{i}+1\equiv \Phi_{i}$  will be the posterior precision after reception of an \textit{i.i.d.} signal. Thus, $\Phi _i >\nu _{i}$. Notice that $\mu_{i}=\frac{\nu_{i}}{\nu_{i+1}}$.}. Therefore, \textit{our theory of confidence predicts the overprecision phenomenon} even before completion of the task. In contrast with the other distortions of confidence, underprecision will never be observed, a prediction which is corroborated by \cite{moore2008} who do not quote any study in their discussion of "underprecision". The overestimation of the precision of acquired knowledge is an additional manifestation of the hidden search undertaken by intuitive Bayesians. Our analysis of overprecision is congruent with the observation that greater overconfidence of this kind was found for tasks in which subjects considered they were more competent \citep{heath1991}. 

\section{Predicting confidence biases and learning} \label{biases_and_learning}

\subsection{Confidence updating by intuitive Bayesians} \label{updating}

In our experiment, confidence is reported prior to the task $E_{1}p$, after four rounds $E_{5}p$, and after nine rounds (only for doublers) $E_{10}p$.

After going through four rounds of anagrams, a number of cues on the task have been received and processed. Participants may recall how many anagrams they solved in each round and in the aggregate, whether they would have passed the test in each round or on the whole at this stage of the task, whether their performance improved or declined from one round to the next, how fast they could solve anagrams, and so forth. For the purpose of decision-making, cues are converted into a discrete set of \textit{i.i.d.} Bernoulli variables taking value 1 if they signal to the individual that she should reach her goal for level $l$ $(l=1,2,3)$, and 0 otherwise. The single parameter of the Bernoulli variable is its mean which defines the expected likelihood of success. However, this mean is essentially unknown to that individual. Thus, let it be denoted by $\widetilde{p}$ which is randomly distributed within the interval [0, 1]. Assume that the prior distribution of $\widetilde{p}$ is a Beta-distribution with a reported mean $E_{1}p$ and precision $\nu$.

Behaving like intuitive Bayesians, participants update their prior expectation of success at level $l$ $(l=1,2,3)$ before the training session $E_{1}p_{l}$ in the following manner (see \citealt{degrooth}, Chapter 9): 

\begin{center}
\begin{equation} \label{eq3}
E_{5}p_{l}=\frac{\nu _{l}}{\nu _{l}+\tau _{4l}}E_{1}p_{l}+\frac{1}{\nu _{l} +\tau _{4l}}X_{1-4;l}
\end{equation}
\end{center}

\noindent
with $\tau_{4l} >0$ designating the precision of all the independent cues perceived during the first four rounds, and $X_{1-4;l}$ defining  the number of independent cues predicting future success at level $l$ at this stage of the task. They also update the precision of the posterior expectation  $E_{5}p_{l}$, which rises from $\nu _{1l}$ to: 
               
\begin{center}
\begin{equation} \label{eq3b}  \tag{3'} \nu _{5l} = \nu _{1l} +\tau _{4l} 
\end{equation}
\end{center}

\noindent
with $0\leq X_{1-4,l}\leq \tau _{4l}$.

Equation (\ref{eq3}) cannot be directly estimated on the data because the estimated probabilities  $E_{1}p_{l}$ and  $E_{5}p_{l}$ are unobservable. However, it may be rewritten concisely in terms of reported confidence $q_{1} (l)$ and $q_{5} (l)$ with the help of the miscalibration equations (\ref{eq1}) and (\ref{eq2}). Let us express generally the Bayesian transformation of the probability estimates into confidence as:

\begin{center}
\begin{equation} \label{eq4}
q_{5}(l)=\mu _{5l}E_{5}p_{l}+(1-\mu _{5l})D_{5,l}\, ,\: \: l=(1,2,3)
\end{equation}
\end{center}

\noindent
with $\mu _{5l}=\frac{\nu _{l}+\tau _{4l}}{\nu _{l}+\tau _{4l}+1}$ and\footnote{In order to have an unambiguous definition of $D_{(5,l)}$ and $D_{(1,l)}$ below, we use the expected utility (EU) criterion, as explained in note \ref{rationaldecision}.} 

\noindent
$$
D_{(5,l)} = \left\{
    \begin{array}{ll}
       1 & \mbox{if } max\, EU(l'\mid E_{5} p_{l'},l'=(0,\cdots l-1))\geq max\, EU({l}''\mid E_{5}p_{{l}''},{l}''=(l,\cdots, 3)) \\
        0 & \mbox{otherwise.}
    \end{array}
\right.
$$

Confidence is merely a weighted average of the prior forecast and a \textit{doubt term} acting as a contrarian Bernoulli signal. 

And likewise:
\begin{center}
\begin{equation} \label{eq5}
q_{1}(l)=\mu _{1l}E_{1}p_{l}+(1-\mu _{1l})D_{1,l}
\end{equation}
\end{center}

\noindent
with $\mu _{1l}=\frac{\nu _{l}}{\nu _{l}+1}$ and 

\noindent
$$
D_{(1,l)} = \left\{
    \begin{array}{ll}
       1 & \mbox{if } max\, EU(l'\mid E_{1} p_{l'},l'=(0,\cdots l-1))\geq max\, EU({l}''\mid E_{1}p_{{l}''},{l}''=(l,\cdots, 3)) \\
        0 & \mbox{otherwise.}
    \end{array}
\right.
$$
\\
\\
Combining (\ref{eq3}), (\ref{eq4}) and (\ref{eq5}), we get:

\begin{center}
\begin{equation} \label{eq6}
q_{5}(l)=\frac{\nu _{l}+1}{\nu _{l}+\tau _{4l}+1}q_{1}(l)+\frac{1}{\nu _{l}+\tau _{4l}+1}X_{1-4,l}+\frac{1}{\nu _{l}+\tau _{4l}+1}(D_{5,l}-D_{1,l})
\end{equation}
\end{center}

By the same reasoning, we can express the confidence of doublers for upper levels $l=(2,3)$ as:

\begin{center}
\begin{equation} \label{eq7}
q_{10}(l)=\frac{\nu _{l}+\tau _{4l}+1}{\nu _{l}+\tau _{9l}+1}q_{5}(l)+\frac{1}{\nu _{l}+\tau _{9l}+1}X_{5-9,l}+\frac{1}{\nu _{l}+\tau _{9l}+1}(D_{9,l}-D_{5,l})
\end{equation}
\end{center}

\noindent
with $\tau _{9l}\geq \tau _{4l}$ designating the precision of all of the independent cues perceived during the training level (9 rounds), $\nu _{l}+\tau _{9l}$ the precision of the posterior expectation $E_{10}p_{l}$, and $X$ defining the number of independent cues predicting future success at level $l$ at this stage of the task. 

Equations (\ref{eq6}) and (\ref{eq7}) are essentially the same with a moving prior of increasing precision. In the absence of miscalibration, confidence reported before round $t (t=(5,10))$ would be a weighted average of prior confidence and the mean frequency of cues predicting future success at level $l$ since the last time confidence was reported. With miscalibration, another term is added which can only take three values, reflecting the occurrence and direction of change in subjects' estimated ability with experience. If experience confirms the prior intention to stop or continue to level $l$, this additional term takes value 0 and confidence is predicted by the rational-Bayesian model (with perfect calibration). However, if experience disconfirms the prior intention to stop or continue to level $l$, confidence rises above this reference value with disappointing experience and declines symmetrically below this reference value with encouraging experience.\textit{ Thus, our model predicts that intuitive Bayesians be conservative} and under-react symmetrically to negative experience (by diminishing their confidence less than they should) and to positive experience (by raising their confidence less than they should). Below, we report indeed rather small variations of confidence in our experiment in the form of local, but not global, learning.

\subsection{Regression analysis}\label{regression_section}

The models of Bayesian estimation of confidence described by equations (\ref{eq6}) and (\ref{eq7}) are tested by an OLS in Tables \ref{bayes1} and \ref{bayes2}\footnote{The discrete value of confidence between 0 and 100 can be safely treated as continuous.} respectively. 
Reported confidence in participant $i$'s ability to reach one level of the double-or-quits game is regressed in Table \ref{bayes1} (Table \ref{bayes2}) on the confidence that she reported before the first (fifth) round and on a vector $Z_{li}$ of level-specific cues observable in the first four (last five) rounds, assuming that $X_{1-4,li}(X_{5-9,li})=\beta _{l}Z_{li}+\epsilon _{li}$ where $\beta _{l}$ is a vector of coefficients and $\epsilon _{li}$ an error term of zero mean. Two dummy variables for the hill and choice treatments (wall as reference) have been added to the regression.

\begin{table}[!ht]
\begin{center}
\caption{\label{bayes1} \small{OLS estimation of the Bayesian model of confidence before round 5}}

% Table generated by Excel2LaTeX from sheet 'table1'
\footnotesize

\begin{tabular}{rccc}

\hline
           & {\bf Training Level } & {\bf Middle Level } & {\bf High Level } \\
\hline
Confidence before training session &   $ 0.79^{***} $ &   $ 0.86^{***} $ &   $ 0.90^{***} $ \\

Freq. of rounds with 4 anagrams solved &   $ 0.14^{***} $ &    $ 0.06^{ns} $ &    $ 0.01^{ns} $ \\

Freq. of rounds with 5- 6 anagrams  solved &   $ 0.29^{***} $ &   $ 0.19^{***} $ &   $ 0.13^{***} $ \\

Freq. of rounds with non-declining performance  &   $ 0.12^{***} $ &   $ 0.10^{***} $ &   $ 0.09^{***} $ \\

Anagrams solved per minute on rounds 1-4 &   $ 0.01^{***} $ &   $ 0.01^{***} $ &   $ 0.01^{***} $ \\

      Hill &     $ 0.03^{*} $ &    $ 0.04^{**} $ &     $ 0.03^{*} $ \\

    Choice &    $ 0.01^{ns} $ &    $ 0.02^{ns} $ &    $ 0.00^{ns} $ \\

  Constant &  $ -0.25^{***} $ &  $ -0.25^{***} $ &  $ -0.20^{***} $ \\
\hline
        $ R^{2} $ &       67\% &       70\% &       76\% \\
        Observations&410&410&410\\
\hline\hline
\multicolumn{4}{l}{
 \begin{minipage}{1\textwidth}%
 \medskip
    \scriptsize Notes. \textbf{Significance level}: * \(p<.10\), ** \(p<.05\), *** \(p<.01\), ns: not significant at 10\% level. \textbf{Variables}: Frequency of rounds with non-declining performance represents the percentage of rounds (in rounds 2-4) in which number of anagrams solved was equal or higher than in the previous round, it takes four values (0,.33,.67,1). Hill and Choice: dummy variables with Wall as reference.
  \end{minipage}
  } %}\\
\end{tabular}
\end{center}
\end{table}

\begin{table}[!ht]
\begin{center}
\caption{\label{bayes2} \small{OLS estimation of the Bayesian model of confidence for doublers reported before the middle level}}
\normalsize
% Table generated by Excel2LaTeX from sheet 'table1'
\footnotesize
\begin{tabular}{rcc}
\hline
           & {\bf Middle Level } & {\bf High Level } \\
\hline
Confidence after round 4 &  $  0.772^{***} $ &   $ 0.872^{***} $ \\

Freq. of rounds with 4 anagrams solved (5-9) &    $ 0.017^{ns} $ &   $ -0.024^{ns} $ \\

Freq. of rounds with 5- 6 anagrams solved (5-9) &  $  0.120^{***} $ &     $ 0.073^{*} $ \\

Freq. of rounds with non-declining performance (5-9) &    $ 0.034^{ns} $ &    $ 0.088^{**} $ \\

Number of rounds used to solve 36 anagrams &   $ 0.027^{***} $ &    $ 0.021^{**} $ \\

Anagrams solved per minute on rounds 5-9 &   $  0.003^{ns} $ &   $ -0.003^{ns} $ \\

      Hill &  $ -0.047^{***} $ &   $ -0.022^{ns} $ \\

    Choice &   $ -0.017^{ns} $ &    $ 0.022^{ns} $ \\

  Constant &   $ -0.136^{ns} $ &   $ -0.186^{**} $ \\
\hline
       $  R^{2} $ &       74\% &       81\% \\
       Observations& 275&275\\
\hline\hline
\multicolumn{3}{l}{
 \begin{minipage}{1\textwidth}%
 \medskip
    \scriptsize Notes. \textbf{Significance level}: * \(p<.10\), ** \(p<.05\), *** \(p<.01\), ns: not significant at 10\% level. \textbf{Variables}: Frequency of rounds with non-declining performance represents the percentage of rounds (in rounds 5-9) in which number of anagrams solved was equal or higher than in the previous round. Hill and Choice: dummy variables with Wall as reference. Number of rounds used to solve 36 anagrams (between rounds 6 and 9). (5-9) refers to measures between rounds 5 and 9. 
  \end{minipage}
  } %}\\
\end{tabular}
\end{center}
\end{table}

The regressions confirm the existence of local learning. Subjects did revise their expectations with experience of the task as several cues have highly significant coefficients (at 1\% level) with the right sign. Moreover, they analyze their own performance correctly by setting stronger pre-requisites for themselves when the task gets more difficult. For example, their ability to solve just four anagrams per round in the training period increases their confidence for this period only because, if such performance is enough to ensure success in this period, it is no longer sufficient when the task becomes more difficult. Another interesting result in Table 5 consistent with the miscalibration term in equation (\ref{eq7}) concerns low achievers who double. The later they ended up solving the required number of anagrams in the training period, the more abruptly their confidence rose. It is indeed an implication of subjects' vulnerability to illusory signals that low-ability doublers find themselves almost as confident as high-ability doublers in spite of widely different performances. This result appears too on Figures \ref{middle} and \ref{high}, where the ability-adjusted confidence of low-ability doublers jumps from bottom to top during the second stage of the training period.

A major testable implication of the Bayesian model lies in the coefficient of the prior confidence, which must be interpreted as the precision of prior information relative to the information collected by experience of the task during the training period. This coefficient is always high in Tables \ref{bayes1} and \ref{bayes2} with a minimum value of 0.77. Observing such high weights for the prior favors the hypothesis of rational-Bayesian updating over adaptive expectations as the latter would considerably underweight the prior relative to the evidence accumulated in the first four rounds. Successful experience of the easier task in the early rounds is expected to be more predictive of final success on the same task than in future tasks of greater difficulty. Thus, the relative weight of experience should diminish in the confidence equation at increasing levels or, equivalently, the relative weight of prior confidence should rise. Indeed, the coefficient of prior confidence increases continuously with the level. It rises from 0.79 to 0.86 and 0.90 in Table \ref{bayes1}; and, from 0.77 to 0.87 in Table \ref{bayes2}.  In parallel, the coefficients of cues signaling a successful experience continuously diminish when the level rises. We can use the mathematical expressions of the two coefficients of prior confidence derived from equations (\ref{eq6}) and (\ref{eq7}) to calculate the precision of early experience relative to prior confidence (before the task) $\frac{ \tau _{4l}}{\nu _{l}}$  $(l=1,2,3)$. With the data of Table \ref{bayes1}, we get 0.266 for the training level, 0.163 for the middle level, and 0.111 for the high level. Similarly, we compute the precision of late experience relative to prior confidence (before the task) $\frac{ \tau _{9l}}{\nu _{l}}$  $(l=2,3)$. With the data of Table \ref{bayes2}, we get 0.506 for middle level and 0.274 for high level. The impact of learning from experience appears to be substantial and with increasing returns. By elimination of $\nu _{l}$, we finally calculate the precision of early experience relative to total experience during the training period $\frac{ \tau _{4l}}{\tau _{9l}}$  $(l=2,3)$. We obtain 0.322 for middle level and 0.405 for high level. The rate of increase of precision resulting from longer experience (from 4 to 9 rounds) $\frac{\tau _{9l}- \tau _{4l}}{\tau _{4l}}$  reaches a considerable 211\% at middle level and 147\% at high level, which forms indirect evidence of the overprecision phenomenon.

\subsection{Why do intuitive Bayesians make wrong (and costly) predictions of performance?} \label{why}

The answer to this important question, and to the related \textit{planning fallacy}, is contained in Table \ref{bias}, which uses the same set of potential predictors to forecast confidence in succeeding the middle level after doubling and ex post chances of success\footnote{We used an OLS to predict probabilities of success so as to make the comparison with confidence transparent. Estimating an OLS instead of a Probit in columns 3 and 4 didn't affect the qualitative conclusions. }: prior confidence, ability, and performance cues observed subsequently (during rounds 5 to 9). The mere comparison of coefficients between the two columns of Table \ref{bias} demonstrates that posterior confidence is based on both objective performance cues and subjective variables, whereas the chances of success are predicted by the objective performance cues and ability only. The latter are the frequencies of rounds with 4 and with 5-6 anagrams solved respectively (effort) and the speed of anagram resolution (ability); and the subjective variables are essentially the prior confidence and the illusory signal given to low achievers by their (lucky) initial success. Remarkably, the number of rounds needed for solving 36 anagrams (varying from 6 to 9), which indicates low achievement and recommends quitting the game at an early stage, acts as an illusory signal with a significantly positive effect on confidence in column 1; but the same variable acts as a correlate of low ability in column 2 with a strong negative effect on the chances of success at middle level. Indeed, the subjective predictors of posterior confidence do not predict success when the objective performance cues are held constant. Prior confidence predicts the posterior confidence that conditions the decision to double\footnote{Conditional on initial success, prior confidence is a good predictor of the future decision to double (regression not shown). This is good news for the quality of confidence reports; and it confirms that subjects behave as intuitive Bayesians who rely on their own subjective estimates of success to make the choice of doubling.} but fails to predict success because it is based on an intuitive reasoning which suffers from systematic biases. Past errors convey to the prior through the aggregation procedure of Bayesian calculus and may add up with further errors caused by the perception of illusory signals. 

\begin{table}[!h]
\begin{center}
\caption{\label{bias} \small{Estimation of posterior confidence (after doubling) and ex post chances of success at the middle level}}

\footnotesize
% Table generated by Excel2LaTeX from sheet 'table1'
\begin{tabular}{rcc}
\hline
    {\bf } & \multicolumn{ 2}{c}{{\bf Level 2  }} \\

           & {\bf Confidence After} & {\bf Chances of success} \\
\hline
Confidence after round 4 &   $ 0.778^{***} $ &    $ 0.034^{ns} $ \\

Freq. of rounds with 4 anagrams solved (5-9) &  $   0.014^{ns} $ &    $ 0.276^{*} $ \\

Freq. of rounds with 5- 6 anagrams solved (5-9) &   $ 0.107^{***} $ &   $ 0.348^{**} $ \\

Freq. of rounds with non-declining performance (5-9) &   $  0.043^{ns} $ &   $ -0.036^{ns} $ \\

Number of rounds used to solve 36 anagrams &   $ 0.024^{***} $ &  $ -0.115^{***} $ \\

   Anagrams solved per minute on rounds 5-9 &   $ 0.009^{ns} $ &   $ 0.070^{***} $ \\
   
      Ability &   $ -0.007^{ns} $ &   $ 0.062^{***} $ \\

      Hill &  $ -0.046^{***} $ &    $ 0.097^{ns} $ \\

    Choice &   $ -0.018^{ns} $ &    $ -0.100^{*} $ \\

  Constant &   $ -0.106^{ns} $ &    $  0.598^{*} $ \\
\hline
        $ R^{2} $ &       74\% &       30\% \\
        Observations&275&275\\
\hline\hline
\multicolumn{3}{l}{
 \begin{minipage}{1\textwidth}%
 \medskip
    \scriptsize Notes. \textbf{Sample}: to be comparable, these regressions consider only those who succeeded first level and decided to double to second level. \textbf{Significance level}: * \(p<.10\), ** \(p<.05\), *** \(p<.01\), ns: not significant at 10\% level. \textbf{Variables}: Frequency of rounds with non-declining performance represents the percentage of rounds (in rounds 5-9) in which number of anagrams solved was equal or higher than in the previous round. Hill and Choice: dummy variables with Wall as reference. Number of rounds used to solve 36 anagrams (between rounds 6 and 9). (5-9) refers to measures between rounds 5 and 9. Number of rounds used to solve 36 anagrams (between rounds 6 and 9).
  \end{minipage}
  } %}\\
\end{tabular}
\end{center}
\end{table}

To reinforce our demonstration, we used the regressions listed in Table \ref{bias} to predict normative (based on rational expectations) and subjective (confidence-based) expected values\footnote{The predicted values were computed on regressions containing only the significant variables. We checked that these values stayed close to predictions derived from the regressions listed in Table \ref{bias} which contain non significant variables too.}  and determine the best choice of doubling or quitting prescribed by those alternative models. As expected, the normative model's predictions (based on the true -ex post- probabilities) deviate farther from reality than the subjective model's: 48\% versus 17\% of the time. However, the confidence-based prescriptions have no information value since the rate of failure is the same whether one follows the prescription  (69\%) or not (70\%). By contrast, the normative prescriptions have great value since the rate of failure is 52\% for those who respect them versus 88\% for those who don't. Finally, Table \ref{evaluation_2} divides the sample of doublers in four categories: 47\% are able and calibrated, 12\% are unable and calibrated, 36\% are overconfident and 5\% are underconfident. Rates of failure are markedly different among these categories: 52\% only for the able calibrated, 57\% for the (able) underconfident, 78\% for the unable calibrated and 91\% for the (unable) overconfident! Undeniably, the prevalence of miscalibration among doublers is substantial and its cost in terms of failure is massive.

\begin{table}[!ht]
\begin{center}
\caption{\label{evaluation_2} The prevalence and cost of miscalibration among doublers}
% Table generated by Excel2LaTeX from sheet 'table1'
\footnotesize
\begin{tabular}{ccccc}
\hline\hline
Presciption of subjective&Prescription of normative&&&Rate of\\
 expected value &  expected value& Category &Share& failure \\
\hline\hline
double & double& able and calibrated&47\%&52\%\\
stop & stop& unable and calibrated&12\%&78\%\\
double & stop& overconfident&36\%&91\%\\
stop & double& underconfident&5\%&57\%\\
\hline \hline
\end{tabular}
\end{center}
\end{table}

\section{Conclusion} \label{conclusion}
We designed an experimental analog to the popular double-or-quits game to compare the speed of learning one's ability to perform a task in isolation with the speed of rising confidence as the task gets increasingly difficult.  In simple words, we found that people on average learn to be overconfident faster than they learn their true ability.  We present a new intuitive-Bayesian model of confidence which integrates confidence biases and learning. The distinctive feature of our model of self-confidence is that it rests solely on a Bayesian representation of the cognitive process: intuitive people predict their own probability of performing a task on the basis of cues and contrarian illusory signals related to the task that they perceive sequentially. Confidence biases arise in our opinion, not from an irrationality of the treatment of information, but from the poor quality and subjectivity of the information being treated. For instance, we rule out self-attribution biases, motivated cognition, self-image concerns and manipulation of beliefs but we describe people as being fundamentally uncertain of their future performance and taking all the information they can get with limited discrimination, including cognitive illusions. Above all, a persistent doubt about their true ability is responsible for their perception of contrarian illusory signals that make them believe, either in their possible failure if they should succeed or in their possible success if they should fail.

Our intuitive-Bayesian theory of estimation combines parsimoniously the cognitive bias and the learning approach. It brings a novel interpretation of the cognitive bias and it provides a general account of estimation biases. Indeed, we did not attribute confidence biases to specific cognitive errors but to the fundamental uncertainty about one's true ability; and we predicted phenomena beyond the hard-easy and Dunning-Kruger effect which could not be explained all together by previous models: miscalibration and overprecision \textit{before} completion of the task, limited discrimination, conservatism, slow learning and planning fallacy. Moreover, we showed that these biases are likely to persist since the Bayesian aggregation of past information consolidates the accumulation of errors, and the perception of illusory signals generates conservatism and under-reaction to events. Taken together, these two features may explain why intuitive Bayesians make systematically wrong and costly predictions of their own performance.  Don't we systematically underestimate the time needed to perform a new (difficult) task and never seem to learn?

Our analysis of overconfidence is restricted to the overestimation bias. The latter must be carefully distinguished from the overplacement bias since the hard-easy effect that we observed here with absolute confidence has often been reversed when observing relative confidence: overplacement for an easy task (like driving one's car) and underplacement for a novel or difficult task. The reasons for overplacement are probably not unique and context-dependent. When people really compete, the over (under) placement bias may result from their observing and knowing their own ability (although imperfectly) better than others'. If both high-ability and low-ability individuals compare themselves with average-ability others, the former are likely to experience overplacement and the latter underplacement. The same reasoning applies to individuals familiar or unfamiliar with the task, and to individuals who were initially successful or unsuccessful with the task. When no real competition is involved, the overplacement effect relates to an evaluation-based estimate of probability. While there is an underlying choice to be made in the estimation task, no such thing is present in the latter case. If I ask you whether you consider yourself as a top driver (relative to others), I don't generally expect you to show me how you drive. Preference reversals are not uncommon between choices and evaluations \citep{lichtenstein1971reversals}. Thus, the present analysis of overestimation is consistent with reasonable explanations of overplacement. Moreover, it predicts the overprecision phenomenon and even rules out underprecision. This demonstrates that overestimation and overprecision are related but different biases. 

Double-or-quits-type behavior can be found in many important decisions like addictive gambling \citep{goodie2005}, military conquests \citep{Johnson2004}, business expansion \citep{malmendier2005}, speculative behavior \citep{shiller2000}, educational choices \citep{breen2001}, etc. Overconfident players, chiefs, entrepreneurs, traders, or students are inclined to take excessive risks; they are unable to stop at the right time and eventually fail more than well-calibrated persons\footnote{However, overconfidence may pay off when there is uncertainty about opponents' real strengths, and when the benefits of the prize at stake are sufficiently larger than the costs (e.g., \citealt{johnson2011, anderson2012}). }  (e.g., \citealt{barber2001, camerer1999}). In contrast, under-confident individuals won't take enough risks and stay permanently out of successful endeavors. 

On the theoretical side, the intuitive-Bayesian model of confidence before completion of a task creates a link between confidence and decision analyses and their respective biases. Confidence biases and the anomalies of decision under risk or uncertainty can be analyzed with the same tools. The estimation of one's ability implies an implicit comparison between an uncertain binary lottery and a reference outcome. It is a by-product of the question: should I double or quit? This is a question of interest to behavioral and decision theorists.\\

\begin{acknowledgements}
We thank the French \textit{Minist\'ere de la Recherche} (\textit{ACI "Contextes sociaux, contextes institutionnels et rendements des syst\`emes \'educatifs"}) for generous support, Claude Montmarquette for offering an opportunity to conduct part of the experimental sessions at CIRANO (Montreal), and Noemi Berlin for numerous discussions. We are grateful to the referees and the editors of this special issue for bringing very helpful remarks and suggestions. We remain responsible for any error.

\end{acknowledgements}

\bibliographystyle{spbasic}      % basic style, author-year citations
\bibliography{references}   % name your BibTeX data base

\clearpage
\section*{Appendix} \label{appendix4}
\renewcommand{\thesection}{\Alph{section}}
\setcounter{section}{0} 
%\appendix

\begin{figure}[h!]
\begin{center}
\caption{Example of the task screen}
\label{screen}
\begin{minipage}{1\textwidth} % choose width suitably
\includegraphics[width=\linewidth]{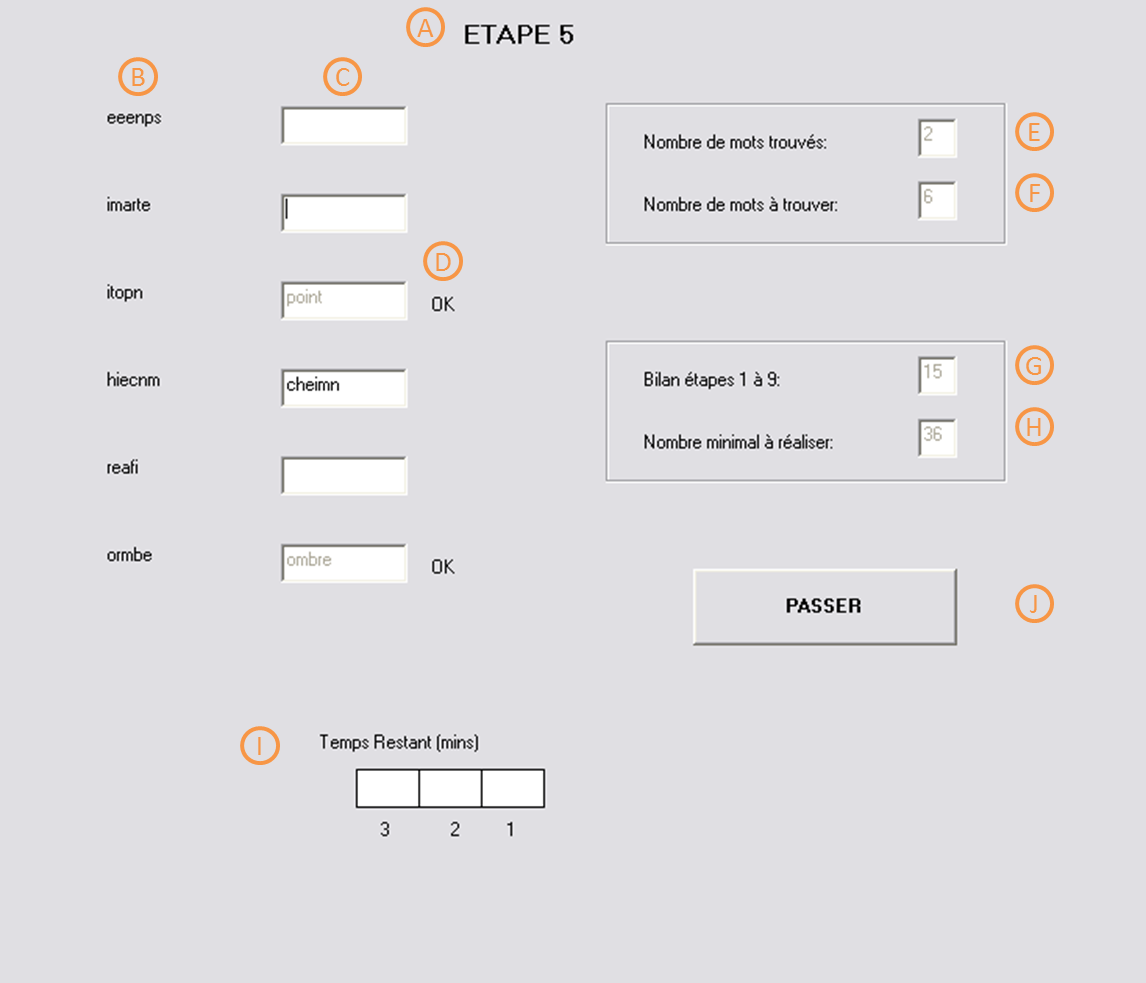}
{\footnotesize Legend:\\
A: Actual round (round 5 in this example).\\
B: List of anagrams to be decoded.\\
C: Fields to type the correct word.\\
D: Feedback. The "OK" appears when the solution for the anagram is correct.\\
E: Number of correct anagrams in the current round.\\
F: Total anagrams to be decoded in the current round, 6 in this example (first level).\\
G: Number of cumulated correct anagrams, including the current and previous rounds.\\
H: Number of correct anagrams required to solve the current level, in this example 36 (first level).\\
I: Remaining time. The total time is 8 minutes, we show only the 3 last minutes.\\
J: Button to go to next round. Participants can pass to next round without clearing all anagrams in the current level, but they cannot come back once they pushed the button.
\par}
\medskip 
\end{minipage}
\end{center}
\end{figure}

\end{document}